\documentclass[9pt,conference]{IEEEtran}
\usepackage{amssymb,amsthm,amsmath,array}
\usepackage{graphicx}
\usepackage[caption=false,font=footnotesize]{subfig}
\usepackage{xspace}
\usepackage[sort&compress, numbers]{natbib}
\usepackage{stmaryrd}
\usepackage{xcolor}
\usepackage{mathtools}
\usepackage{float}
\usepackage{textcomp}

\begin{document}
\title{Magnetic Phase Imaging \\ using Lorentz Near-field Electron Ptychography}
\author{\IEEEauthorblockN{\small
        Shengbo You\IEEEauthorrefmark{1},
        Peng-Han Lu\IEEEauthorrefmark{2},
        András Kovács\IEEEauthorrefmark{2},
        Thomas Schachinger\IEEEauthorrefmark{3,4},
        Frederick Allars\IEEEauthorrefmark{1},
        Rafal E. Dunin-Borkowski\IEEEauthorrefmark{2}, and 
        Andrew M. Maiden\IEEEauthorrefmark{5}
    }
    \IEEEauthorblockA{\small
        \IEEEauthorrefmark{1} Department of Electronic and Electrical Engineering, University of Sheffield, Sheffield S1 4DE, United Kingdom.\\
        \IEEEauthorrefmark{2} Ernst Ruska-Centre for Microscopy and Spectroscopy with Electrons and Peter Grünberg Institute, Forschungszentrum Juelich, 52425 Juelich, Germany.\\
        \IEEEauthorrefmark{3} Institute of Solid State Physics, TU Wien, 1040 Wien, Austria.\\
        \IEEEauthorrefmark{4} University Service Centre for Transmission Electron Microscopy, TU Wien, 1040 Wien, Austria.\\
        \IEEEauthorrefmark{5} Diamond Light Source, Diamond House, Harwell Science and Innovation Campus, Didcot OX11 0DE, United Kingdom.}
}
\maketitle
\begin{abstract}
    Over the past few years, the combination of diffuser and near-field electron ptychography has drawn more attention by its ability to recover large field of view with few diffraction patterns. In this paper, we purpose a novel design and implementation of amplitude diffuser. The amplitude diffuser introduces structures to the illumination while reducing the inelastic scattering. And the amplitude diffuser is implemented at the condenser lens aperture, allowing us to vary the illumination size under the same microscope setup. We demonstrate the reconstruction results under both conventional Transmission Electron Microscopy (TEM) mode as well as Lorentz mode. 
\end{abstract}

\section{Introduction}
Ptychography is a phase-retrieval approach that is becoming widely used in electron microscopy. Ptychography is a form of scanned-probe microscopy, in which a convergent electron beam – the ptychographic probe – illuminates a small patch of a sample, and then scans through a grid of positions to cover a region of interest, recording diffraction patterns at each point in the grid. The diffraction patterns are then iteratively processed to reconstruct a complex-valued transmission function of the sample, whilst simultaneously recovering the probe wavefront \cite{R1}. This simple experimental process and relatively robust reconstruction algorithms have led to ptychography’s uptake across a range of wavelength regimes, from light \cite{R2}, to X-rays \cite{R3}\cite{R4} and electrons \cite{R5}\cite{R6}\cite{R7}\cite{R8}\cite{R9}\cite{R10}\cite{R11}\cite{R12}.

However, due to the near-focused probe and the requirement for significant overlap of the scan positions, covering a large field of view using ptychography can require a large number of diffraction patterns, often numbering in the tens of thousands. Near-field electron ptychography, on the other hand, is capable of overcoming this limitation. Near-field electron ptychography replaces the focused electron beam with a full-field structured illumination and moves the detector from a far-field into a near-field diffraction condition. Near-field ptychography was first demonstrated with X-rays \cite{R13} but has since been extended to visible light \cite{R14} and electron \cite{R15} microscopy. In previous implementations of near-field electron ptychography \cite{R15}\cite{R16}, the sample was illuminated by a broad, roughly parallel illumination beam, and the selected area aperture (SAA) used to mask off a region of the resulting bright field image. An engineered phase plate (or diffuser) inserted in the SAA \cite{R15} modulating the wavefront exiting the SAA plane, before it propagated a short distance to a secondary image plane where it was recorded by a detector. Compared to far-field electron ptychography, this setup could reconstruct a large field of view with as few as nine diffraction patterns. However, the sample was constantly exposed to the electron beam, and changing the magnification of the setup to image at different resolutions involved changing the size of the SAA, so that a full set of diffuser-equipped apertures were required to operate over a set of magnifications. The phase plates also introduced additional unwanted inelastic scatter into the recorded diffraction data. In this talk, we overcome these limitations, whilst retaining near-field ptychography’s merits, through a new experimental arrangement in which the diffuser is located in the condenser aperture rather than the SAA.   

\begin{figure}[t]
    \centering
    \includegraphics[scale = 0.33]{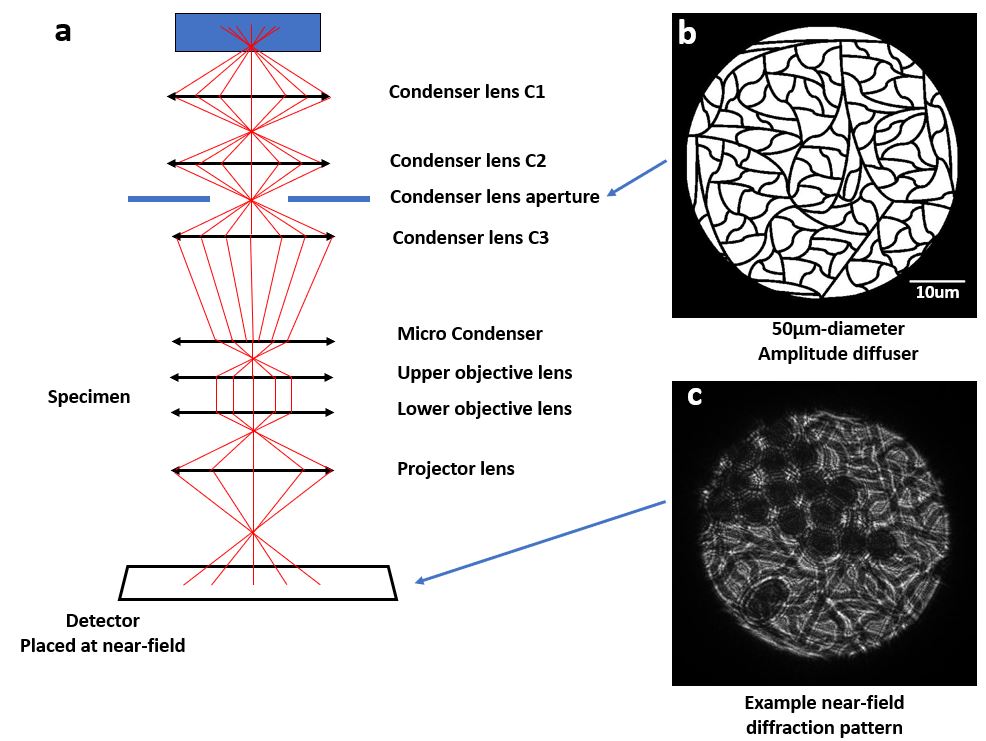}
    \caption{a) the microscope setup. The amplitude diffuser is inserted at the condenser lens aperture. The specimen is illuminated by the broad parallel beam with structures from the amplitude diffuser. b) is the 50µm-diameter amplitude diffuser. c) is an example near-field diffraction pattern. Note that both the amplitude diffuser and the sample are in defocus state.}
    \label{fig:my_label}
\end{figure}
Our microscope configuration, developed from previous work with electron and visible-light microscopes \cite{R14}\cite{R15}\cite{R16}, is shown in Figure 1a. The main difference over previous implementations is that an amplitude-only diffuser, rather than a phase plate, is inserted in the condenser aperture rather than at a downstream aperture plane (i.e., in the SAA or objective aperture). The amplitude diffuser is a 50 µm-diameter etched Silicon Nitride window, with randomly curved lines across the entire diffuser as shown in Fig 1b. The amplitude-only diffuser avoids the introduction of inelastic scattering by a phase plate. Moving the diffuser from the selected area aperture (SAA) up to the condenser aperture means there is no masking of the beam by apertures further downstream, which is an inefficient use of dose; it means the illumination size is no longer limited by the physical size of SAA and instead, various illumination sizes can be programmed by adjusting the condenser lens settings; and it means the method can be expanded to operation in a field-free (Lorentz) mode for imaging of magnetic fields, as we will show in Figure 2. The microscope is aligned and operated in conventional TEM bright field image mode, but with a substantial defocus. This produces near-field diffraction patterns similar to that shown in Figure 1c. 

A significant difference compared to previous implementations of near-field electron ptychography \cite{R15} is the use of an amplitude diffuser instead of a phase diffuser. With the phase diffuser, a Silicon Nitride membrane with a random thickness profile introduces variable phase delays into the electron beam, to produce a randomly structured illumination profile. However, this introduces additional unwanted inelastic scatter, which forms a strong contrast-reducing background signal to the diffraction data. The amplitude diffuser is intended to similarly allow structure to be introduced to the illumination, whilst reducing the additional inelastic background in the recorded data. The amplitude diffuser is composed of 350nm wide randomly arranged tracks within a 50 µm aperture. The tracks completely block the electron beam, such that the slightly defocused image of the condenser aperture incident on the sample contains a pattern of fringes and phase distortions. 

Provided sufficient structure is introduced by the fringes in the illumination, ptychographic reconstruction can quantitatively phase imaging for the specimen. By reducing the inelastic scattering using amplitude diffuser, together with separating the inelastic scattering background using the ePIE algorithm, the reconstructed magnitude images show better contrast than previous near-field electron ptychography experiments \cite{R15}\cite{R16}.

\begin{figure}[t]
    \centering
    \includegraphics[scale = 0.33]{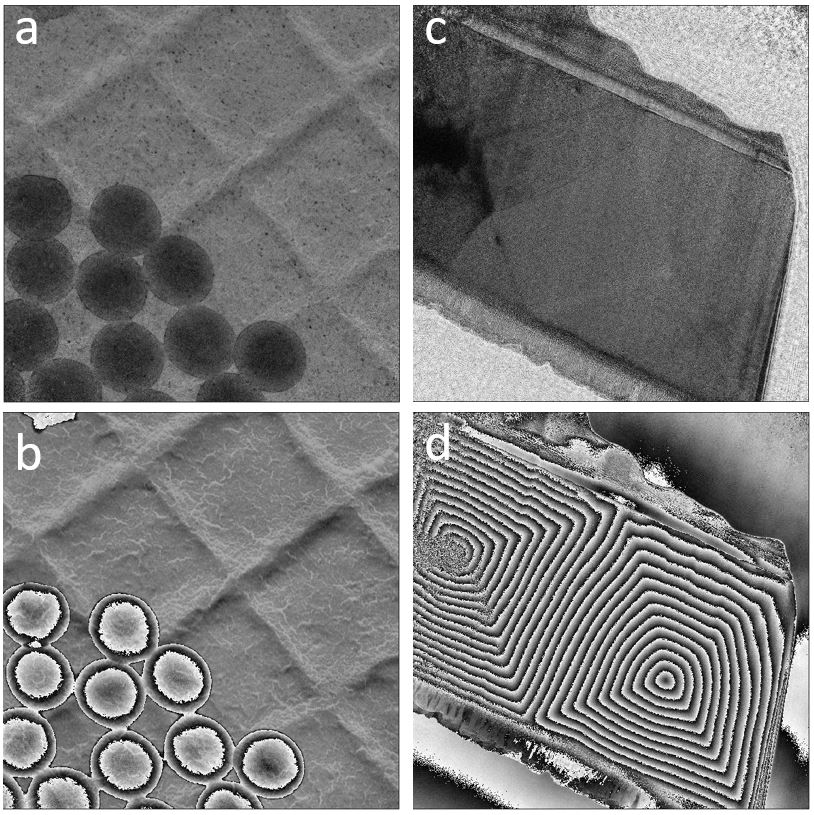}
    \caption{Latex sphere and Mo-doped Permalloy sampled reconstructed results under conventional TEM mode and Lorentz mode respectively. (a) and (b) are the reconstructed amplitude and phase of latex sphere sample under conventional TEM mode. (c) and (d) are the reconstructed amplitude and phase of Mo-doped Permalloy. As for the magnetic sample. }
    \label{fig:my_label}
\end{figure}

\section{Conclusion}
In this talk, we show experimental results that near-field electron ptychography can be used to reconstruct magnetic sample under Lorentz mode with various illumination sizes. We move the diffuser from the selected area aperture to the condenser lens. This improves the does efficiency and allows us to use different illumination sizes, while
keeping the merits such as large field of view with few diffraction patterns. The purposely-designed amplitude diffuser reduce inelastic scattering and makes the illumination more coherent. A parallel beam of electrons illuminates onto the specimen. A
system diagram is shown in Figure 1, together with a picture of the amplitude diffuser and an example of diffraction pattern. The reconstructed sample results are shown in Figure 2.

\bibliographystyle{IEEEtran}
\bibliography{references}

\end{document}